# Bell Correlations From Local Un-entangled States of Light and Quantum Electro-dynamics

Louis Sica[1,2]

[1]Institute for Quantum Studies, Chapman University, Orange, CA & Burtonsville, MD, USA
[2]Inspire Institute Inc., Alexandria, VA, USA
Email:lousica@jhu.edu

Based on the Bell theorem, it has been believed that a theoretical computation of the Bell correlation requires explicit use of an entangled state. Such a physical superposition of light waves occurs in the down-converter sources used in Bell experiments. However, this physical superposition is eliminated by wave propagation to spatially separated detectors. Bell correlations must therefore result from local waves, and the source boundary conditions of their previously entangled state. In the present model, Bell correlations are computed from disentangled separated waves, boundary conditions of nonlinear optics, and properties of single-photon and vacuum states specified by quantum electrodynamics. Transient interference is assumed between photon-excited waves and photon–empty waves based on the possibility of such interference found to be necessary by the designers of Bell-experiment sources. The present model employs local random variables without specifying underlying causality.

## 1. Introduction

The Bell theorem and violation of Bell's inequalities by experimentally acquired data sets have been believed to make the derivation of Bell correlations impossible without the existence of a perpetual entanglement based non-locality. However, it is shown in [1] that the mere existence of three or four data sets of $\pm 1's$, whether their origin is in experimental observation or counterfactual prediction, implies cross correlations that identically satisfy the corresponding Bell inequality independently of Bell's assumptions, and even of whether the data are random. The cross correlations among data sets may have varying functional forms, except for special processes such as wide sense stationarity. In the quantum mechanical case, conditional probabilities due to non-commutation of measurements beyond one per particle produce sets of correlations that are different from those assumed by Bell. The claim that experimental data violate the Bell inequality is commonly based on the use of six or eight data sets obtained in statistically independent pairs instead of the three or four cross-correlated sets used in the original derivation of the corresponding Bell inequality. Unsurprising, the inequality is violated. (See [1] for a detailed discussion of the issues summarized above.)

As stated, Bell's assumptions in deriving the Bell inequality are unnecessary: it holds under general mathematical conditions without them. Of course, nullifying the Bell theorem does not automatically imply that a local model for the Bell correlation in the absence of entanglement is possible. However, superimposed wave pairs originating in spontaneous-parametric-down-conversion (SPDC) crystals become physically separated by experimental design, implying that Bell correlations can occur among photons attached to un-entangled local wave pairs due to physical source boundary conditions.

The current paper provides a calculation of the Bell correlation that begins with the principle that physically separated, non-superimposed, electromagnetic waves do not instantaneously influence each other. (The notion that they do is based on the flawed interpretation of Bell experiments stated above.) The correlations are computed from waves originating in nonlinear interactions, and standard interpretations of quantum electro-dynamic (QED) photon-wave properties. The originally entangled waves, some containing photons, others photon empty, later transiently interfere, consistent with the design [2,3] of Bell-experiment sources. Further assumptions used in the model are that light consists of photons and waves, and that single photons do not divide at beam splitters, although interference still occurs at interferometer outputs (see Jacque, et. al. [4]).

In QED theory, the quantum states of electromagnetic waves consist of ground/vacuum states and excited states that are interpreted as photons. The excited states are thus assumed herein to consist of photons attached to waves while the ground state is interpreted as a photon-empty wave. A recent classic experiment by Jacque, et. al. [4] suggests that photon and associated vacuum states may interfere, and that interpretation is adopted in the present model. It is also consistent with the stated design of Bell experiments, for which it was found necessary to enable interference between photon wave states and vacuum wave states by equalizing optical paths. Thus modern experiments suggest that light is a composite of waves and attached photons rather than an entity that switches from particle-like to wave-like behavior depending on experimental circumstances.

However, the fact that little is actually known about photons complicates efforts to develop a local physical model for the Bell correlation dependent on initial conditions at photon creation. In addition, properties of the quantum vacuum state are not well known, as its magnitude as given by QED has been called into serious question. (See Meis [5] for enumeration of various quantitative discrepancies and their discussion.) Thus, in the following, use of the QED values for photon-containing and photon-empty waves is to be viewed as a specific hypothesis that leads to a derivation of the Bell correlation.

Two pairs of orthogonally polarized waves (see Figure), one photon attached to each pair, are emitted by an SPDC source with an added path equalization component (see source design in [2,3]). Interference between polarization components of a photon-containing wave and an accompanying photon-empty wave occurs in coordinate systems rotated with respect to the original SPDC source system of coordinates. The phases of waves are assumed to be statistically independent of their amplitudes as is also consistent with the phase uncertainty of single photon states. Phase matching conditions of SPDC together with energy conservation constrain phase sums, but allow phase differences that may fluctuate over successive photon and vacuum waves. No mechanism is specified for the association of photons with waves beyond the formalism of QED [6], nonlinear optics, and experimental observation. Interactions are assumed to be local but variables' local behavior is random without attempts at further hidden variable explanations rooted in detailed photonic behavior.

The formalism of this paper begins in a way that is parallel to that used in [7]. However, the physical and probability models evolve differently. Local wave properties of light as employed here have also been used in [8] to account for polarization correlations of entangled photons after assuming circular polarization emanating from the source.

## 2. Bell Correlations from Local Pairs of Photons

Based on the geometry of type II SPDC sources for Bell correlation experiments [2,3], two amplitudes $\vec{U}_1$ and $\vec{U}_2$ are introduced representing Beams 1 and 2 of the Figure. These are vector amplitudes that result from super-positions of orthogonal vector components $u_{iH}, u_{iV}, i=1,2$. Thus,

$$\vec{U}_1 = u_{1H}\hat{i} + u_{1V}\hat{j} \tag{2.1a}$$

and

$$\vec{U}_2 = u_{2H}\hat{i} + u_{2V}\hat{j}. \tag{2.1b}$$

Initial horizontal and vertical amplitudes are indicated by subscripts H and V that correspond to unit vectors $\hat{i}$ and $\hat{j}$ respectively. The H and V components are both present in two secondary source regions labeled Beam 1 and Beam 2 in the figure as used in Bell experiments. (For simplicity, finite beam effects such as diffraction are neglected in the following model.)

The components are given more explicitly as:

$$\begin{aligned}u_{1H} &= |u_{1H}|\cos(\theta_{1H} + \omega_{1H}t + 2\pi x/\lambda_{1H}); & u_{1V} &= |u_{1V}|\cos(\theta_{1V} + \omega_{1V}t + 2\pi x/\lambda_{1V})\\ u_{2H} &= |u_{2H}|\cos(\theta_{2H} + \omega_{2H}t + 2\pi x/\lambda_{2H}); & u_{2V} &= |u_{2V}|\cos(\theta_{2V} + \omega_{2V}t + 2\pi x/\lambda_{2V}),\end{aligned} \tag{2.2}$$

where subscript pairs 1H–2V and 1V–2H correspond to phase-matched conservation-of-energy linked wave pairs in the down-converter source. Coordinate x is measured from the output plane of the source to the four detectors, all assumed to be equidistant from the source.

The experimental requirements of nonlinear optics (phase matching and conservation of energy) are assumed so that pairs of photons are emitted at random, one photon per amplitude pair in each of Beams 1 and 2. Each beam thus has a photon containing amplitude component and an orthogonally polarized photon-empty component. From the requirements of SPDC type II, each emitted photon pair has either polarizations 1H-2V or 1V-2H, each occurring with probability one half, where the numerals indicate the beams into which the photons are deposited. The laser pump intensity is adjusted so that two such events, i.e. four photons, rarely occur simultaneously.

Four QED-ground-state waves to which two photons per emission event become attached are assumed to be initially present in the crystal. The SPDC crystals are configured to produce wave-pair components 1H2V and 1V2H that separate to become Beams 1 and 2, and propagate to individual polarization analyzer-detectors on sides A and B of Bell experiments. Equal optical paths between the source and the four detectors are assumed. In the source crystal, phase matching occurs for each wave pair due to the symmetry of the source crystal structure [2,3]. A compensator crystal, rotated $90^0$ with respect to the first, results in all the beams having traversed equal optical paths after exiting the assembly so that orthogonal component pairs in Beams 1 and 2 may exhibit transient interference in analyzer outputs.

Due to the experimental design used, the polarization beam splitters on sides A and B of a Bell experiment will be illuminated by beams having random polarizations. The polarization components in the transmit-reflect directions will be linear combinations of the components of the orthogonal pairs in Beams 1 and 2 indicated above.

The action of polarization analyzers placed in each of Beams 1 and 2 is now computed. Transmit and reflect components of $\vec{U}_1$ and $\vec{U}_2$ depend on the angle of rotation of the analyzer with respect to the $\hat{i}$ direction. Orthogonal unit vectors in the (n) and (p) directions of the two analyzers are

$$\hat{n}_{ln} = \cos\theta_l \hat{i} + \sin\theta_l \hat{j}, \quad \hat{n}_{lp} = -\sin\theta_l \hat{i} + \cos\theta_l \hat{j} \quad l=1,2. \tag{2.3}$$

From these, one obtains the analyzer's transmitted and reflected output amplitudes based on inputs (2.1) and (2.2):

$$\begin{aligned} U_{ln} &\equiv \vec{U}_l \cdot \hat{n}_{ln} = u_{lH}\cos\theta_l + u_{lV}\sin\theta_l, \\ U_{lp} &\equiv \vec{U}_l \cdot \hat{n}_{lp} = -u_{lH}\sin\theta_l + u_{lV}\cos\theta_l. \end{aligned} \quad l=1,2 \tag{2.4a}$$

The instantaneous intensities of the analyzer output components n and p for each of the inputs $\vec{U}_1$ and $\vec{U}_2$ are given by

$$I_{ln} = U_{ln}^2 \; ; \; I_{lp} = U_{lp}^2 \quad l=1,2 \tag{2.4b}$$

where $I_{1n}$ is explicitly evaluated using (2.2) to clarify the notation:

$$\begin{aligned} I_{1n} = &|u_{1H}|^2 \cos^2(\theta_{1H} + \omega_{1H}t + 2\pi x/\lambda_{1H})\cos^2\theta_1 + |u_{1V}|^2 \cos^2(\theta_{1V} + \omega_{1V}t + 2\pi x/\lambda_{1V})\sin^2\theta_1 + \\ &|u_{1H}||u_{1V}|\cos(\theta_{1H} + \omega_{1H}t + 2\pi x/\lambda_{1H})\cos(\theta_{1V} + \omega_{1V}t + 2\pi x/\lambda_{1V})\sin 2\theta_1. \end{aligned} \tag{2.5}$$

This can be simplified by performing short time averages over terms that oscillate at optical frequency: $\cos(\theta + \omega t)$. These fast oscillations play no role in the model calculation. Thus Eq. (2.5) becomes:

$$\begin{aligned} I_{1n} =& \frac{|u_{1H}|^2 \cos^2\theta_1}{2} + \frac{|u_{1V}|^2 \sin^2\theta_1}{2} + \\ & \frac{|u_{1H}||u_{1V}|\cos[(\theta_{1H} - \theta_{1V}) + (\omega_{1H} - \omega_{1V})t + 2\pi x(1/\lambda_{1H} - 1/\lambda_{1V})]\sin 2\theta_1}{2} \\ =& I_{1H}\cos^2\theta_1 + I_{1V}\sin^2\theta_1 + \sqrt{I_{1H}I_{1V}}\cos[(\theta_{1H} - \theta_{1V}) + (\omega_{1H} - \omega_{1V})t + 2\pi x(1/\lambda_{1H} - 1/\lambda_{1V})]\sin 2\theta_1 \end{aligned} \tag{2.6}$$

where absolute-value-squared amplitudes divided by 2 are replaced by equivalent average intensities and the only retained term dependent on time occurs at a lower beat frequency. In a similar manner

$$I_{1p} = I_{1H}\sin^2\theta_1 + I_{1V}\cos^2\theta_1 - \sqrt{I_{1H}I_{1V}}\cos[(\theta_{1H}-\theta_{1V})+(\omega_{1H}-\omega_{1V})t+2\pi x(1/\lambda_{1H}-(1/\lambda_{1V})]\sin 2\theta_1 \quad (2.7)$$

The original source intensities at $\theta_1 = 0$ are then $I_{1n}(0) = I_{1H}$ and $I_{1p}(0) = I_{1V}$. For analyzer 2 one obtains

$$I_{2n} = I_{2H}\cos^2\theta_2 + I_{2V}\sin^2\theta_2 + \sqrt{I_{2H}I_{2V}}\cos[(\theta_{2H}-\theta_{2V})+(\omega_{2H}-\omega_{2V})t+2\pi x(1/\lambda_{2H}-1/\lambda_{2V}]\sin 2\theta_2 \quad (2.8)$$

and

$$I_{2p} = I_{2H}\sin^2\theta_2 + I_{2V}\cos^2\theta_2 - \sqrt{I_{2H}I_{2V}}\cos[(\theta_{2H}-\theta_{2V})+(\omega_{2H}-\omega_{2V})t+2\pi x(1/\lambda_{2H}-1/\lambda_{2V}]\sin 2\theta_2 \quad (2.9)$$

At $\theta_2 = 0$, $I_{2n}(0) = I_{2H}$ and $I_{2p}(0) = I_{2V}$.

The equations may be simplified by using specifications of nonlinear optics [9] in SPDC-phase matching and symmetry of the source:

$$\begin{aligned}\theta_{2H} + \theta_{1V} &= const + \Delta_{2H},\\ \theta_{2V} + \theta_{1H} &= const + \Delta_{2V},\end{aligned} \quad (2.10a)$$

where the $\Delta$'s are additional phase shifts implemented by a wave plate used in experiments [2,3]. The difference of phases in the two beams is then

$$\theta_{2H} - \theta_{2V} = \theta_{1H} - \theta_{1V} + \Delta_{2H} - \Delta_{2V}. \quad (2.10b)$$

The condition $\Delta_{2H} - \Delta_{2V} = \pi$, is experimentally implemented so that

$$\theta_{2H} - \theta_{2V} = \theta_{1H} - \theta_{1V} + \pi \quad (2.10c)$$

In addition, angular oscillation frequencies of (2.6-2.9) are related through conservation of energy relations in the nonlinear source:

$$\omega_{1H} + \omega_{2V} = \omega_p; \quad \omega_{1V} + \omega_{2H} = \omega_p \quad (2.10d)$$

where $\omega_p$ equals the angular frequency of the pump from which the pairs of photons are derived in the nonlinear process. From the relations of Eq. (2.10d) it follows that

$$\omega_{1H} - \omega_{1V} = \omega_{2H} - \omega_{2V}: \quad 1/\lambda_{1H} - 1/\lambda_{1V} = 1/\lambda_{2H} - 1/\lambda_{2V}, \quad (2.10e)$$

where the maximum relative frequency variation between the beams has been estimated as [3] $d\omega/\omega = .007$. While conditions (2.10a-e) are physically reasonable, a separate experimental study of source properties would be necessary to confirm them in detail.

When (2.10c) and (2.10e) are used in (2.6-2.9), one obtains

$$\begin{aligned}I_{1n} &= I_{1H}\cos^2\theta_1 + I_{1V}\sin^2\theta_1 + \sqrt{I_{1H}I_{1V}}\cos[(\theta_{1H}-\theta_{1V})+(\omega_{1H}-\omega_{1V})t+2\pi x(1/\lambda_{1H}-1/\lambda_{1V})]\sin 2\theta_1\\ I_{1p} &= I_{1H}\sin^2\theta_1 + I_{1V}\cos^2\theta_1 - \sqrt{I_{1H}I_{1V}}\cos[(\theta_{1H}-\theta_{1V})+(\omega_{1H}-\omega_{1V})t+2\pi x(1/\lambda_{1H}-1/\lambda_{1V})]\sin 2\theta_1\\ I_{2n} &= I_{2H}\cos^2\theta_2 + I_{2V}\sin^2\theta_2 - \sqrt{I_{2V}I_{2H}}\cos[(\theta_{1H}-\theta_{1V})+(\omega_{1H}-\omega_{1V})t+2\pi x(1/\lambda_{1H}-1/\lambda_{1V})]\sin 2\theta_2\\ I_{2p} &= I_{2H}\sin^2\theta_2 + I_{2V}\cos^2\theta_2 + \sqrt{I_{2V}I_{2H}}\cos[(\theta_{1H}-\theta_{1V})+(\omega_{1H}-\omega_{1V})t+2\pi x(1/\lambda_{1H}-1/\lambda_{1V})]\sin 2\theta_2.\end{aligned} \quad (2.11ad)$$

Since the phase differences are not determined by Equations (2.10a-c,e), the value of $\cos[(\theta_{1H}-\theta_{1V})+(\omega_{1H}-\omega_{1V})t+2\pi x(1/\lambda_{1H}-1/\lambda_{1H})]$ varies randomly over the interval +1 to -1 depending on the precise phases, frequency, wavelength of the light, and photon pulse occurrence time, and averages to zero over random photon occurrence times, while $\cos^2[(\theta_{1H}-\theta_{1V})+(\omega_{1H}-\omega_{1V})t+2\pi x(1/\lambda_{1H}-1/\lambda_{1V})]$ averages to ½. The relative phases among the waves as they exit the source are maintained at equal times/distances from the source since they all propagate with the same velocity. Note however, that if $\omega_{1H}=\omega_{1V}$ and $\lambda_{1H}=\lambda_{1V}$, i.e., the original entangled waves have the same wavelength and frequency, that the detectors can be at different distances from the source without affecting the results to be derived below.

. For the physical situation considered, the two equations of (2.10a) correspond alternately to photon-containing wave pairs and vacuum wave pairs. It is not clear whether phase behavior differences should be attributed to waves depending on whether they do or do not contain photons. However, to obtain agreement with correlations that result from physical entanglement and enable wave interference, the phase conditions that hold for photon-containing waves will be assumed to hold for photon-empty waves.

The individual intensity variables (2.11a-d) are now interpreted from QED to clarify their later use in computation of the Bell correlation. The use of QED concepts will be illustrated for variables $I_{1n}$ and $I_{1p}$ where these wave-pulse intensities will be defined as proportional to the QED photon energies when counts occur. Computations will be carried out for two alternative sets of relative intensities based on the QED [6] description of a single photon state:

$$I_{1H}=I_{2V}=1 \text{ for a photon wave-pulse, } I_{1V}=I_{2H}=1/2 \text{ for vacuum waves,}$$

or

$$I_{1V}=I_{2H}=1 \text{ for a photon wave-pulse, } I_{1H}=I_{2V}=1/2 \text{ for vacuum waves,}$$

each condition occurring with probability ½ over photon-pair emission events, and with energies/intensities given in units of $h\nu$ by QED for photon pulses in observation time windows. Angle $[\theta_{1H}-\theta_{1V}+(w_{1H}-\omega_{1V})t+2\pi x(1/\lambda_{1N}-1/\lambda_{1V})]$ is replaced by $\theta$, and as stated above, $\overline{\cos\theta}=0$ over randomly occurring beats evaluated at photon occurrence times.

Since beam intensities are measured by photon occurrences, the intensities for photon pair production events 1H2V and 1V2H are equal: $I_{1H}=I_{2V}$ and $I_{1V}=I_{2H}$ if the less than 1% frequency variation among the relevant beams is neglected. Beam intensity pairs without photon occurrences are assumed to be equal also, and in accordance with QED to have intensity ½. The corresponding beam amplitudes are given by the square roots of the intensities. (Overbars and rectangular brackets are used interchangeably to denote averages below.)

The mean of $I_{1n}$ in Eq. (2.11a) under the above source specification is given by

$$\overline{I_{1n}} = \frac{1}{2}\left(1\cos^2\theta_1 + (1/2)\sin^2\theta_1 + 1\cdot(1/\sqrt{2})\overline{\cos\theta}\sin 2\theta_1\right)_{1H}$$
$$+ \frac{1}{2}\left((1/2)\cos^2\theta_1 + 1\sin^2\theta_1 + 1\cdot(1/\sqrt{2})\overline{\cos\theta}\sin 2\theta_1\right)_{1V}$$
(2.12a)

where subscripts on the parentheses indicate which input variable is associated with a photon. Similarly, the mean of $I_{1p}$ from Eq. (2.11b) is

$$\overline{I_{1p}} = \frac{1}{2}\left(1\sin^2\theta_1 + (1/2)\cos^2\theta_1 - 1\cdot(1/\sqrt{2})\overline{\cos\theta}\sin 2\theta_1\right)_{1H}$$
$$+ \frac{1}{2}\left((1/2)\sin^2\theta_1 + 1\cos^2\theta_1 - 1\cdot(1/\sqrt{2})\overline{\cos\theta}\sin 2\theta_1\right)_{1V}$$
(2.12b)

Equations (2.12a,b) become

$$\overline{I_{1n}} = \frac{1}{2}\left(1\cos^2\theta_1\right)_{1H} + \frac{1}{2}\left(1\sin^2\theta_1\right)_{1V} = \frac{1}{2}$$
(2.12c)

and

$$\overline{I_{1p}} = \frac{1}{2}\left(1\sin^2\theta_1\right)_{1H} + \frac{1}{2}\left(1\cos^2\theta_1\right)_{1V} = \frac{1}{2},$$
(2.12d)

after deleting terms with coefficient ½, and setting $\overline{\cos\theta}$ equal to 0 over averaged photon events and times. Terms with coefficient ½ correspond to pure vacuum contributions to intensities $I_{1n}$ and $I_{1p}$ and are invisible to detectors. They also have different frequencies from the photon-carrying waves. The $\cos\theta$ term alternately increases and decreases probabilities of $I_{1n}$ and $I_{1p}$, but averages to zero over multiple photon counts.

The correlation of $I_{1n}$ and $I_{1p}$ is now computed for photon occurrences 1H and 1V, that occur with probability ½, respectively:

$$\langle I_{1n}I_{1p}\rangle_{1H} = \left\langle \left(1\cos^2\theta_1 + (1/2)\sin^2\theta_1 + 1\cdot(1/\sqrt{2})\cos\theta\sin 2\theta_1\right)_{1H} \times \left(1\sin^2\theta_1 + (1/2)\cos^2\theta_1 - 1\cdot(1/\sqrt{2})\cos\theta\sin 2\theta_1\right)_{1H}\right\rangle$$
$$= 1\cdot 1\cos^2\theta_1\sin^2\theta_1 + 1\cdot(1/2)\cos^4\theta_1 + (1/2)\cdot 1\sin^4\theta_1 + (1/2)^2\sin^2\theta_1\cos^2\theta_1 - (1/2)\overline{\cos^2\theta}\sin^2 2\theta_1$$
$$= 1\cdot 1\cos^2\theta_1\sin^2\theta_1 - (1/2)^2\sin^2 2\theta_1 = 0,$$
(2.13a)

where first power terms in $\overline{\cos\theta}$ have been averaged to zero over multiple events. Similarly,

$$\langle I_{1n}I_{1p}\rangle_{1V} = \left\langle \left((1/2)\cos^2\theta_1 + 1\sin^2\theta_1 + 1\cdot(1/\sqrt{2})\cos\theta\sin 2\theta_1\right)_{1v} \times \left((1/2)\sin^2\theta_1 + 1\cos^2\theta_1 - 1\cdot(1/\sqrt{2})\cos\theta\sin 2\theta_1\right)_{1v}\right\rangle$$
$$= (1/4)\cos^2\theta_1\sin^2\theta_1 + 1\cdot(1/2)\cos^4\theta_1 + (1/2)\cdot 1\sin^4\theta_1 + 1\cdot 1\sin^2\theta_1\cos^2\theta_1 - (1/2)\overline{\cos^2\theta}\sin^2 2\theta_1$$
$$= 1\cdot 1\sin^2\theta_1\cos^2\theta_1 - (1/2)^2\sin^2 2\theta_1 = 0.$$
(2.13b)

The evaluation of Equations (2.13a,b) has been carried out so as to be consistent with the response of two separated detectors, each assumed to have an efficiency of 1 for detection of photons but to be blind to vacuum waves. As a result, terms having a coefficient consisting of a single or no 1's multiplied by factors of ½ are dropped since they correspond to the possibility of activation of only one detector or none. The term with two 1's in the coefficient, multiplied by the probabilities of detection by alternate detectors, is canceled by the interference term. Thus,

$$\langle I_{1n} I_{1p} \rangle = \frac{1}{2} \langle I_{1n} I_{1p} \rangle_{1H} + \frac{1}{2} \langle I_{1n} I_{1p} \rangle_{1V} = 0, \qquad (2.14)$$

consistent with the requirement that photons arrive at either detector 1n or 1p, resulting in a count correlation of zero. It should be noted that the interference terms in Equations (2.11a) and (2.11b) are of opposite sign, as are the resulting fluctuations of intensities $I_{1n}$ and $I_{1p}$ that ultimately specify the average count rates at detectors $1n$ and $1p$. Since each detector then receives an average of half the photons created, the average of $I_{1n} + I_{1p}$ equals 1.

The assumption (indicated by experimental observation) that vacuum waves may interfere with photon bearing waves so as to affect the probability of photon detection yields reasonable results in the above calculations. It should be noted that the different interpretations of photon and non-photon terms follow from the experimental result that photons are not divided at beam splitters while wave intensities are divided. Thus for a photon containing wave, although sine and cosine squared terms correspond to wave division, they also indicate probabilities of photon deflection, but for a vacuum wave they are interpreted as wave intensity division only.

The same interpretation of the photon pair production process used above may be used on the other equations of (2.11a-d). From the photon probability variables given in (2.11 a-d), one may compute the joint intensity or photon count correlations in terms of products such as $I_{1n} I_{2p}$. As above, since $[\theta_{1H} - \theta_{1V} + (w_{1H} - \omega_{1V})t + 2\pi x(1/\lambda_{1H} - 1/\lambda_{1V}]$ is assumed to vary over $2\pi$ given random phases and times of photon occurrence, $\overline{\cos\theta} = 0$, with its square equal to ½ in product terms. Starting from (2.11a,d), the multi-event correlation is

$$\langle I_{1n} I_{2p} \rangle = \langle I_{1H} I_{2H} \cos^2\theta_1 \sin^2\theta_2 + I_{1V} I_{2V} \sin^2\theta_1 \cos^2\theta_2 + I_{1H} I_{2V} \cos^2\theta_1 \cos^2\theta_2 \\ + I_{1V} I_{2H} \sin^2\theta_1 \sin^2\theta_2 + \sqrt{I_{1H} I_{1V}} \sqrt{I_{2H} I_{2V}} \overline{\cos^2\theta} \sin 2\theta_1 \sin 2\theta_2 \rangle, \qquad (2.15)$$

where averaging has already been applied to cross-terms involving $\cos\theta$ in (2.15).

Further evaluation of (2.15) will be shown in detail to illustrate the use of the SPDC generation of photon pairs under the QED model: $I_{1H} = I_{2V} = 1$, $I_{1V} = I_{2H} = 1/2$ or $I_{1V} = I_{2H} = 1$, $I_{1H} = I_{2V} = 1/2$, each occurring with probability ½. Then

$$\overline{I_{1n} I_{2p}} = \frac{1}{2}\left[1 \cdot \frac{1}{2}\cos^2\theta_1 \sin^2\theta_2 + \frac{1}{2} \cdot 1 \sin^2\theta_1 \cos^2\theta_2 + 1 \cdot 1\cos^2\theta_1 \cos^2\theta_2 + \frac{1}{2} \cdot \frac{1}{2}\sin^2\theta_1 \sin^2\theta_2 \right. \\ \left. + \sqrt{1 \cdot \tfrac{1}{2}} \sqrt{\tfrac{1}{2} \cdot 1} \frac{1}{2}\sin 2\theta_1 \sin 2\theta_2 \right]_{1H2V} + \frac{1}{2}\left[\frac{1}{2} \cdot 1 \cos^2\theta_1 \sin^2\theta_2 + 1 \cdot \frac{1}{2}\sin^2\theta_1 \cos^2\theta_2 \right. \qquad (2.16) \\ \left. \frac{1}{2} \cdot \frac{1}{2}\cos^2\theta_1 \cos^2\theta_2 + 1 \cdot 1\sin^2\theta_1 \sin^2\theta_2 + \sqrt{\tfrac{1}{2} \cdot 1}\sqrt{1 \cdot \tfrac{1}{2}} \frac{1}{2}\sin 2\theta_1 \sin 2\theta_2 \right]_{1V2H}.$$

The square brackets' subscripts indicate the alternative ways that two photons may occur, each with probability ½. The terms that imply the possibility of one photon being contributed to each of $I_{1n}$ and $I_{2p}$ are those having a product of two 1's in their coefficients. These terms are the third term in the first square bracket, the fourth term in the second bracket and the two interference terms involving square roots, one from each bracket. The contributions from these terms add to

$$\overline{I_{1n}I_{2p}} = \frac{1}{2}\left[\ 1\cdot 1\cos^2\theta_1 \cos^2\theta_2 + 1\cdot 1\sin^2\theta_1 \sin^2\theta_2 + 2\sin\theta_1 \cos\theta_1 \sin\theta_2 \cos\theta_2\ \right] \quad (2.17a)$$
$$= \frac{1}{2}\cos^2(\theta_1 - \theta_2).$$

From similar analysis, one computes the other correlations:

$$\overline{I_{1p}I_{2n}} = \frac{1}{2}\cos^2(\theta_1 - \theta_2); \quad \overline{I_{1n}I_{2n}} = \overline{I_{1p}I_{2p}} = \frac{1}{2}\sin^2(\theta_1 - \theta_2). \quad (2.17b)$$

By attaching minus signs to the 1p2n and 1n2p averages one obtains the same result as computed from entanglement,

$$-\overline{I_{1n}I_{2p}} - \overline{I_{1p}I_{2n}} + \overline{I_{1n}I_{2n}} + \overline{I_{1p}I_{2p}} = -\cos 2(\theta_1 - \theta_2), \quad (2.18)$$

which is the Bell correlation.

It is useful to re-derive (2.18) in an alternate way to illustrate the internal consistency of the method. The definition of functions $S_1(\theta_1)$ and $S_2(\theta_2)$ is

$$S_1(\theta_1) = I_{1n} - I_{1p} = (I_{1H} - I_{1V})\cos 2\theta_1 + 2\sqrt{I_{1H}I_{1V}}\cos\theta\sin 2\theta_1$$
$$S_2(\theta_2) = I_{2n} - I_{2p} = -(I_{2V} - I_{2H})\cos 2\theta_2 - 2\sqrt{I_{2V}I_{2H}}\cos\theta\sin 2\theta_2. \quad (2.19)$$

Computations are again carried out under alternative QED conditions $I_{1H} = I_{2V} = 1$, $I_{1V} = I_{2H} = 1/2$, and $I_{1V} = I_{2H} = 1$, $I_{1H} = I_{2V} = 1/2$, each with probability 1/2. Consistent with this, one obtains

$$\overline{S_1} = \left\langle (I_{1n} - I_{1p}) \right\rangle = \frac{1}{2}\left(1\cos 2\theta_1 + 2\cdot 1\cdot (1/\sqrt{2})\overline{\cos\theta}\sin 2\theta_1\right)_{1H}$$
$$+ \frac{1}{2}\left(-1\cos 2\theta_1 + 2\cdot 1\cdot (1/\sqrt{2})\overline{\cos\theta}\sin 2\theta_1\right)_{1V} = 0, \quad (2.20)$$

since $\overline{\cos\theta} = 0$. Similarly, $\overline{S_2(\theta_2)} = 0$ and $\overline{S_1^2} = \overline{S_2^2} = 1$. One may now calculate the Bell correlation as a multi-event average and replace first power terms involving $\overline{\cos\theta}$ with zeros:

$$\overline{S_1(\theta_1)S_2(\theta_2)} = \left\langle -(I_{1H} - I_{1V})(I_{2V} - I_{2H})\cos 2\theta_1 \cos 2\theta_3 - 4\sqrt{I_{1H}I_{1V}I_{2V}I_{2H}}\overline{\cos^2\theta}\sin 2\theta_1 \sin 2\theta_2 \right\rangle$$
$$= \frac{1}{2}\left[(-1)(1)\cos 2\theta_1 \cos 2\theta_3 - 4\frac{1}{2}\frac{1}{2}\sin 2\theta_1 \sin 2\theta_2\right]_{1H2V} + \frac{1}{2}\left[(1)(-1)\cos 2\theta_1 \cos 2\theta_3 - 4\frac{1}{2}\frac{1}{2}\sin 2\theta_1 \sin 2\theta_2\right]_{1V2H} \quad (2.21)$$
$$= -\cos 2(\theta_1 - \theta_2).$$

Thus, the methodology used to compute the Bell correlation gives consistent results under the assumptions used to derive it: factors corresponding to photon pairs have intensities 1 as specified by QED while interfering vacuum waves have intensities ½. The vacuum waves do not contribute to observed counts except as they interfere with waves corresponding to counts.

## 3. Conclusion

A number of authors have found reason to question the Bell theorem [10]. The theorem is purported to be a proof that no local hidden variable model of the Bell correlation can be constructed. The fact that the Bell inequality must be identically satisfied by any data sets measured experimentally or from counterfactual prediction, invalidates the theorem since it follows that quantum mechanical data, once obtained and cross-correlated cannot violate the inequality. Nevertheless, invalidation of the theorem does not in itself imply that local physical models not explicitly computed from entanglement exist for the correlations.

The present work develops a version of a local probability model for the Bell correlation using explicitly separated un-entangled waves, boundary conditions based on nonlinear optics, QED, and transient interference between quantum vacuum-state waves and waves with attached photons. The model is idealized, but that is consistent with our level of understanding of photons and vacuum waves. The Bell correlation results from averaging over many particle pair events but necessarily differs in some details from the conventional model based on entanglement that assumes two spatially separated measurements performed on a perpetual superposition of four waves. Such a perpetual superposition is inconsistent with electro-magnetic wave propagation as known outside of the fatally flawed Bell theorem.

## Acknowledgment
The author would like to thank Michael Hall for a detailed critique of a previous version of this paper that aided in the current revision. He would also like to thank Joe Foreman for stimulating conversations prompting reconsideration of some issues.

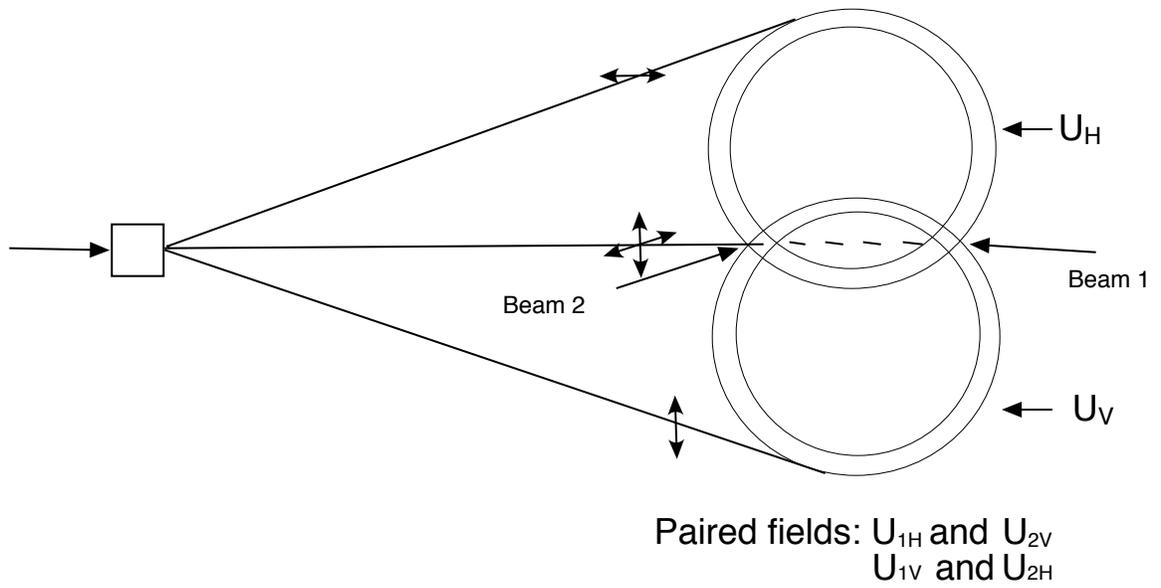

Figure. Output light cones of an SPDC in type II configuration (see Ref. [3]). In experiments, two apertures at the intersections of the light cones indicated result in formation of unpolarized beams 1 and 2 used in Bell experiments.